\documentclass[10pt]{article}
\usepackage[left=4cm,right=3cm,top=1.5cm,bottom=1cm,includeheadfoot]{geometry}
\usepackage[noblocks]{authblk}
\usepackage{fancyhdr}
\usepackage{graphicx}
\usepackage{svg}
\usepackage{lmodern}
\usepackage{csquotes}
\usepackage{amsmath}
\usepackage{amssymb}
\usepackage{bbold}
\usepackage[hidelinks]{hyperref}
\usepackage{orcidlink} % must come after hyperref for options to work
\usepackage[sort&compress]{cleveref} % must come after hyperref
\usepackage{multicol}
\usepackage{lastpage} % for LastPage ref
\usepackage{xcolor}
\usepackage{titlesec}
\usepackage[font=small,labelfont=bf]{caption}
\usepackage{multicol}
\usepackage{lineno}
\usepackage[
    style=phys,
    biblabel=brackets,
    articletitle=true,
    url=false,
    eprint=false,
    doi=false,
    isbn=false,
    maxbibnames=4
]{biblatex}
\AtEveryBibitem{\clearfield{note}}

\titleformat{\section}{\large\bfseries}{\thesection)}{1.4em}{}
\titleformat{\subsection}{\normalsize\bfseries}{\thesubsection)}{0.95em}{}
\titleformat{\subsubsection}{\normalsize}{\thesubsubsection)}{0.55em}{}

\crefname{equation}{eq.\!}{eqs.\!}
\Crefname{equation}{Eq.\!}{Eqs.\!}
\crefname{figure}{fig.\!}{figs.\!}
\Crefname{figure}{Fig.\!}{Figs.\!}

\renewcommand{\vec}[1]{\boldsymbol{#1}} % spatial vectors

\newcommand{\feq}{f^\mathrm{eq}}

\newcommand{\cs}{c_\mathrm{s}}
\newcommand{\ia}{\alpha}
\newcommand{\ib}{\beta}
\newcommand{\ic}{\gamma}

\newcommand{\x}{\vec{x}}

\renewcommand{\c}{\vec{c}}
\renewcommand{\u}{\vec{u}}
\renewcommand{\deg}{^{\circ}}

\usepackage{wasysym}
\title{Rotationally Symmetric Lattice Boltzmann Pseudopotential Methods for Porous Media Applications}
\newcommand{\shorttitle}{B.\,Kellers et al.: Rotationally Symmetric Lattice Boltzmann Pseudopotential Methods}
\date{October 2026}

\author[1,2,*]{\orcidlink{0000-0001-9791-2724}\,Benjamin Kellers}
\author[1,2,3]{\orcidlink{0000-0003-1449-8172}\,Arnulf Latz}
\author[1,2]{\orcidlink{0000-0003-2336-6059}\,Timo Danner}

\affil[1]{%
German Aerospace Center (DLR),
Institute of Engineering Thermodynamics,
Wilhelm-Runge-Str.\,10,
89081 Ulm, Germany}
\affil[2]{%
Helmholtz Institute Ulm Electrochemical Energy Storage (HIU),
Helmholtzstr.\,11,
89081 Ulm, Germany}
\affil[3]{%
Ulm University,
Institute of Electrochemistry,
Albert-Einstein-Allee~47, 
89081 Ulm, Germany}
\begin{document}
\maketitle
\renewcommand{\thefootnote}{\fnsymbol{footnote}}
\footnotetext[1]{Corresponding author, email:
\href{mailto:benjamin.kellers@dlr.de}{benjamin.kellers@dlr.de}}
\renewcommand{\thefootnote}{\arabic{footnote}}

\thispagestyle{empty}
\pagestyle{fancy}

\fancyhead{} % clear all header fields
\fancyhead[L]{\textit{\shorttitle}}
\fancyhead[R]{\thepage{} of \pageref*{LastPage}}
\fancyfoot{} % clear all footer fields

\begin{abstract}
    \noindent Pseudopotential models are a well established approach for multiphase and multicomponent simulations using the lattice Boltzmann method. Although not the most accurate, they are often chosen for ease of implementation and adaptability. A drawback of the models are spurious currents at the interfaces and particularly at solid walls which introduce numerical artifacts. An improved lattice Boltzmann pseudopotential model is presented. The discrete fluid density gradients in the coupling force are computed with rotationally symmetric gradient operators formulated as a convolution. As a result, spurious currents are reduced in settings where solid walls are unaligned with the lattice directions. Additionally, variations of introducing fluid-solid interaction either equally through both fluids or through a single fluid only are compared. Presented model variants are compared using droplets on walls as a common benchmark, analytically solvable capillary rise experiments (Jurin's law), and porous media imbibition as a practical application. Rotationally symmetric operators outperform the classical lattice weights approach in all tests in terms of accuracy, while avoiding extending the lattice to compute higher order gradients. The choice of the fluid-solid interaction variant depends on the problem under consideration and the desired properties of the resulting model.
\end{abstract}

\section{Introduction}
\label{sec:intro}

Multicomponent and multiphase fluid flow is a very active field of research due to the wide range of applications, spanning from electrochemical reactors \cite{Colli2022} over microchip cooling \cite{Huo2026} to biomedical research \cite{Melka2018,Trusov2016}. Despite the differences and in each application, the simulation of porous media flow is a common topic of interest. Particularly, in the field of energy storage and conversion, such as flow batteries \cite{Kok_2016}, fuel cells \cite{Sarkezi2022}, or the electrolyte filling process in lithium-ion batteries \cite{Shodiev2021,Lautenschlaeger2022} multiphase flow in porous media on different length scales determines characteristic properties and in the end performance of the devices.

A popular method for the modeling of such systems is the lattice Boltzmann method (LBM). There exist several methods to simulate multiple phases or components using LBM \cite{Krueger_2017:LBM_Book}, one of which is the pseudopotential method \cite{ShanChen1993,ShanDoolen1995}. In essence, it introduces multiple phases or multiple components through an interaction force term. Adding this to existing LBM code is straight forward. A self-interacting fluid with a suitable choice of pseudopotential depending on the equation of state will result in phase separation. Multiple components are modeled through cross-interaction forces, which are attractive or repulsive for miscible or immiscible fluids, respectively. The method has been studied extensively in the past, see e.g.\ the reviews in Refs.~\cite{Chen2024,Khan2026} and references therein. While many improvements and adaptations have been suggested over the years, e.g.\ Refs.\  \cite{Shan2006,Coelho2021,Sedahmed2022}, several aspects are still not adequately solved. This includes spurious currents emerging from pseudopotential approaches and uncertainty about suitable wall interactions, which require further refinement of simulation approaches.

In this work, we present an interpretation of the pseudopotential method as a regular spatial gradient on discrete density fields. For convenience and comparability we rewrite the theory in the form of discrete convolutions. The resulting gradient kernels will then be optimized for rotational symmetry and the wall interaction adjusted to reduce spurious currents and numerical artifacts. Improvements of our revised method are benchmarked in relevant physical test cases instead of simplified artificial setups. Although we present all the examples in 2D, the theory is also valid for 3D and it is straight forward to construct the corresponding gradient kernels. 

This article is structured as follows. After this short introduction we give a problem description formulating the pseudopotential model as it is often presented in the literature. The formulation is discussed along with shortcomings of the method in \Cref{sec:problem_description}. In \Cref{sec:methodology}, we present the rewritten convolution form of the model, including necessary adjustments that improve its accuracy. Results will be shown in the \Cref{sec:results}. We will discuss the improvements of our method and give a brief outlook in \Cref{sec:discussion}.

\section{Lattice Boltzmann Pseudopotential Model}
\label{sec:problem_description}

In this section we briefly introduce the LBM used in this study. We also discuss the pseudopotential method as it is conventionally introduced in the literature, before highlighting some of its problems. Since the model come with some freedom of choice in the implementation and parametrization. Some of the details for typical implementation our outlined in \Cref{app:B}.

\subsection{Pseudopotential Model for Multicomponent Fluid Flow}

% \subsubsection{LBM with BGK and BB}

The lattice Boltzmann equation is derived from the Boltzmann equation by discretizing time $t$, space $\x$, and particle velocity $\c_i$ where $i$ labels the direction in the velocity discretization. In absence of external forces the discrete Boltzmann equation can be written as
\begin{equation}
    \Delta f_i(t,\x,\c_i) = \Omega_i \Delta t,
\label{eq:LBM}
\end{equation}
where $f_i$ is a particle distribution that describes non-relativistic, non-interacting particles. $\Omega_i$ defines an arbitrary collision operator. Throughout this work, we will use the single-relaxation time collision (BGK) \cite{BGK} which can be modeled using the discrete collision operator \cite{BGK2,BGK3}
\begin{equation}
    \Omega_i = \frac{1}{\tau} 
        \left[ \feq_i(\rho,\u,\c_i) - f_i(t,\x,\c_i) \right].
\label{eq:BGK}
\end{equation}
It evolves $f_i$ exponentially towards its equilibrium distribution $\feq_i$ in a characteristic time scale $\tau$ called relaxation time. $\feq_i$ is a Maxwell-Boltzmann distribution which only implicitly depends on time and space through the macroscopic fields $\rho(t,\x)$ and $\u(t,\x)$, which denote fluid density and macroscopic fluid velocity, respectively. Those observables are computed from the zeroth and first order moments of the distributions $f_i$ and $\feq_i$.

Writing out the left hand side of \Cref{eq:LBM} and applying the collision operator from \Cref{eq:BGK}
\begin{align}
    &f_i(t+\Delta t, \x + \c_i\Delta t, \c_i) - f_i(t,\x,\c_i) 
        = \Omega_i \Delta t
    \\
    \Leftrightarrow\qquad
    &f_i(t+\Delta t, \x + \c_i\Delta t, \c_i)
        = f_i(t,\x,\c_i) 
            + \frac{1}{\tau} 
                \left[ \feq_i(\rho,\u,\c_i) - f_i(t,\x,\c_i) \right]
                \Delta t,
\end{align}
provides an explicit numerical scheme. We note that solving this equation can conceptually be separated into two parts, namely collision and streaming. First, we compute the particle distribution at time $t$ and position $\x$ in their post-collision state, i.e.\ 
\begin{equation}
    f_i^\star(t,\x,\c_i) = f_i(t,\x,\c_i) + \Omega_i \Delta t.
\end{equation}
After that, we update the time and stream the distribution to their new positions in space via
\begin{equation}
    f_i(t+\Delta t, \x + \c_i\Delta t, \c_i) = f_i^\star(t,\x,\c_i).
\end{equation}
At solid boundaries we apply a bounce-back operator, which macroscopically corresponds to $\u=\vec{0}$ at solid walls, i.e.\ a no-slip
boundary condition. For simplicity, we use a full-way bounce-back scheme in which the distribution is streamed into the solid lattice nodes, where we apply the collision
\begin{equation}
    f_i^\star(t,\x,\c_i) = f_{\bar{i}}(t,\x,\c_{\bar{i}}),
\end{equation}
with the label $\bar{i}$ corresponding to the velocity direction opposite to $i$, such that $\c_{\bar{i}} = -\c_i$ for the whole velocity set.

For implementation reasons the solid nodes will include values for $f_i$ when using full-way bounce-back. However, for the calculation of fluid density it is set to zero within wall lattice nodes, such that 
 \begin{equation}
 \rho(\x) = 
 \begin{cases}
        0, \qquad\text{if solid is present at }\x\\
        \sum_i f_i, \qquad\text{otherwise}.
 \end{cases}
 \end{equation}
Note, that this does not affect the value of particle populations $f_i$ ensuring mass and momentum conservation.

The model is extended to multiple components by considering one distribution $f_i^\kappa$ per fluid component $\kappa$, and coupling them by introducing an interaction force \cite{ShanChen1993,ShanDoolen1995,Shan2006,Shan2008,Krueger_2017:LBM_Book}
\begin{equation}
    \vec{F}_\kappa(t,\x) = - \rho_\kappa(t,\x) 
        \sum_{\kappa'} G_{\kappa\kappa'} \vec{\nabla} \rho_{\kappa'}(t,\x),
\label{eq:force_fluid}
\end{equation}
where $G_{\kappa\kappa'}$ is the strength of the interaction between components $\kappa$ and $\kappa'$. In this work, we assume no self interaction is present, such that $G_{\kappa\kappa}=0$. Additionally, the components will be immiscible, i.e.\, $G_{\kappa\kappa'} > 0$. Note, that we are strictly working in the discretized spaces such that operators like $\vec{\nabla}$ have to be understood as discretized variants of their continuous counterparts.

%%%%%%%%%%%%%%%%%%%%%%%%%%%%%%%%%%%%%%%%%%%%%%%%%%%%%%%%%%%%%%%%%%%%%%%%%%%%%%%
\begin{figure}
    \centering
    \includegraphics[width=0.40\linewidth]{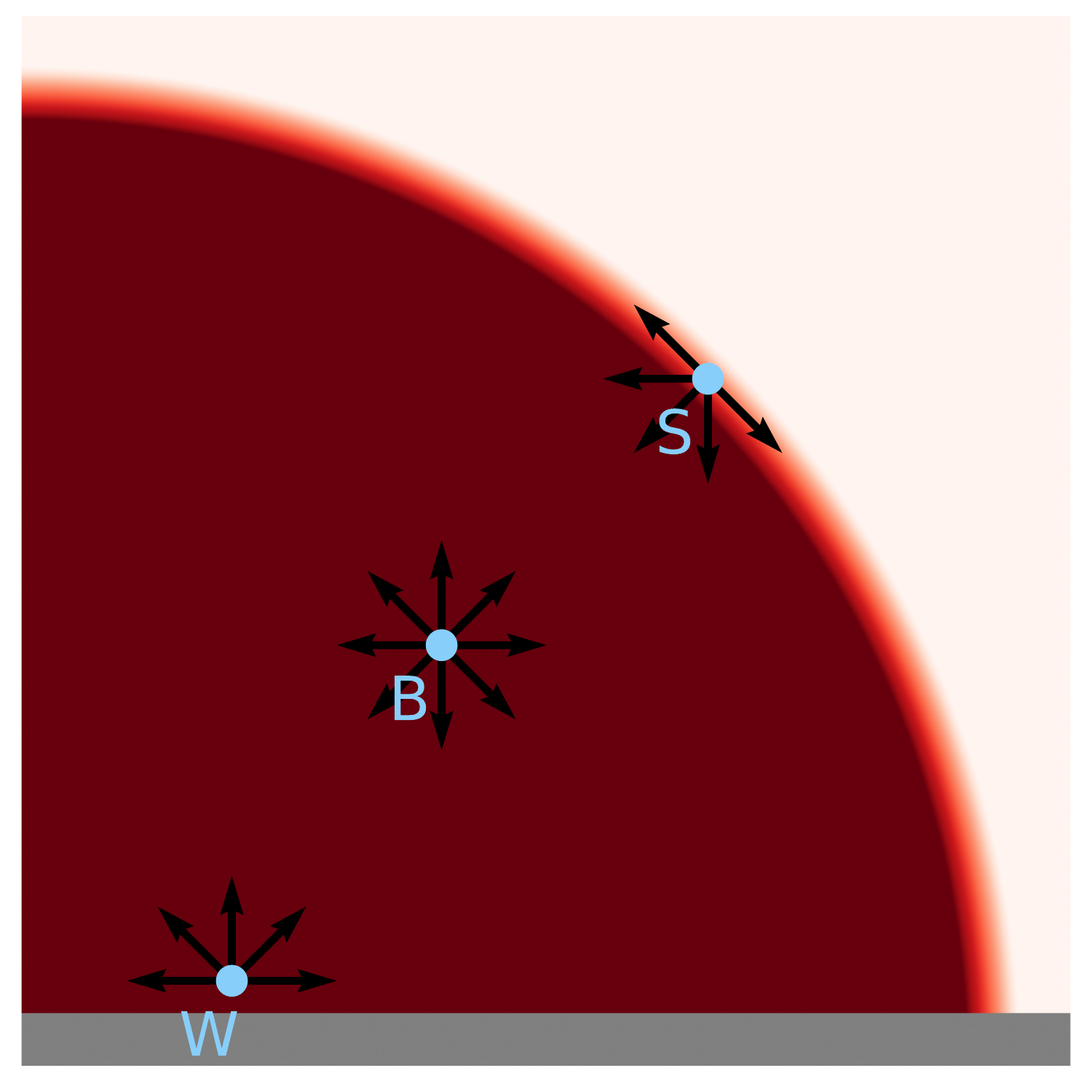}
    \includegraphics[width=0.59\linewidth]{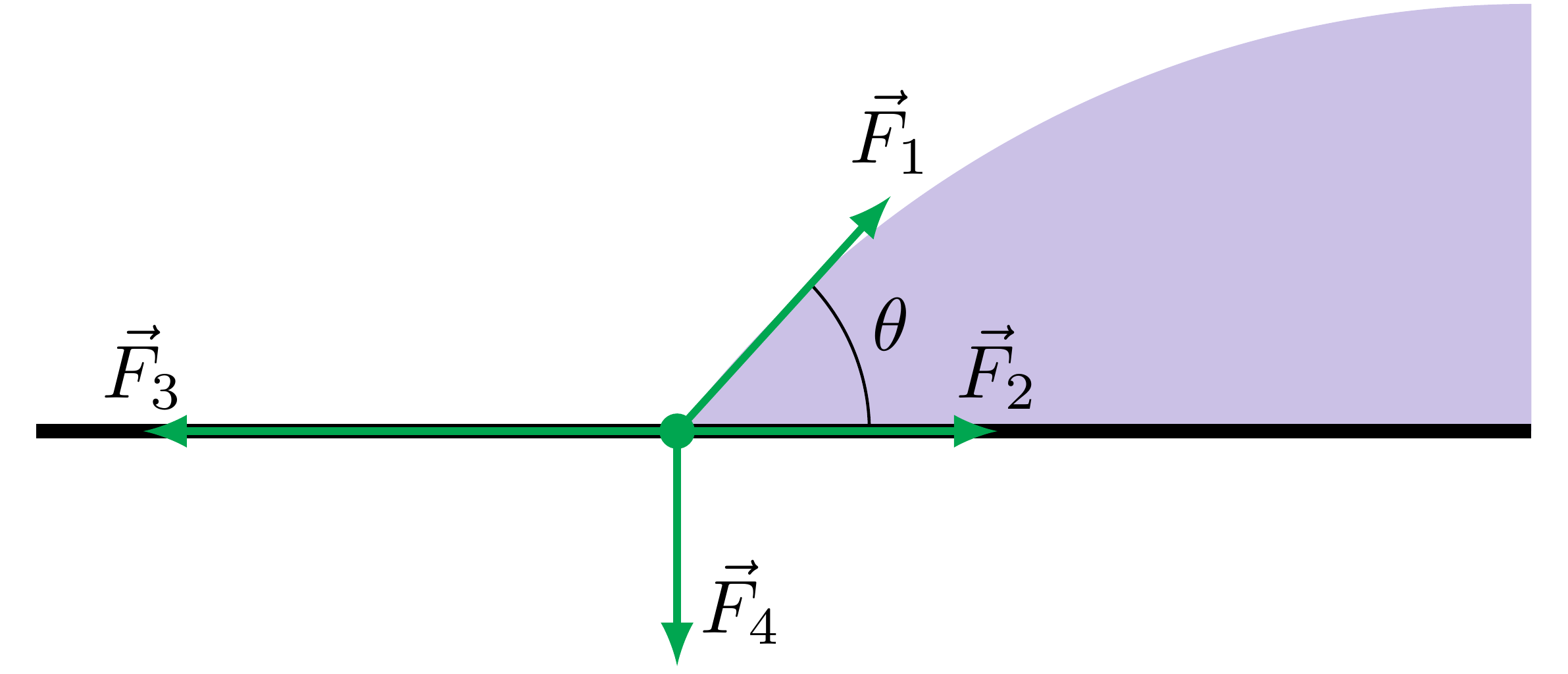}
    \caption{Left: Schematic representation of molecular forces in a droplet of fluid for molecules in the bulk (B), and molecules close to the surface (S) or walls (W). Right: Surface tension ($\vec{F}_1$ to $\vec{F}_3$) forces and solid attraction ($\vec{F}_4$) at the triple point region where two fluid-solid and the fluid-fluid interfaces meet. The contact angle $\theta$ will be such that the total force nets zero. Details in text.}
    \label{fig:forces}
\end{figure}
%%%%%%%%%%%%%%%%%%%%%%%%%%%%%%%%%%%%%%%%%%%%%%%%%%%%%%%%%%%%%%%%%%%%%%%%%%%%%%%

Forces proportional to gradients of density fields tend to result in radial forces pointing towards a center of mass. The can be seen from the schematic representation of forces in \Cref{fig:forces} (left). In the bulk of a fluid bubble, the forces acting upon a molecule from neighboring molecules cancel out since they are evenly distributed around the molecule. This is not the case for molecules close to a solid boundary or to the interface. Here, a net force will emerge that pulls the molecules close to surface towards the center of mass of the bubble. If we now where to compute the gradient of the density field, we would realize that it points exactly in the same direction.

Adding the forces coming from the solid into this picture, we can see why contact angles emerge. \Cref{fig:forces} (right) shows the forces at the triple point where the two fluid-solid and the fluid-fluid interfaces meet. The behavior described through the left picture emerges as surface tension forces, $\vec{F}_1$ to $\vec{F}_3$. Additionally, we consider the molecular attraction from the molecules in the solid wall, $\vec{F}_4$. All four forces net to zero, otherwise an acceleration will be caused that shifts the fluid-fluid surface, until the net force is zero again. This is the reason why a certain contact angle $\theta$ forms where fluid-fluid interfaces are in contact with solid boundaries. 

Basing the force for components $\kappa$ not on its own density field but rather on the density gradients of all the other components $\kappa'\neq\kappa$ leads to component separation. Hence, fluid bubbles will form, which can be considered a simple effect of surface tension. Since we want to study systems of two immiscible fluid components, this is exactly what we want to achieve. \Cref{eq:force_fluid} models cohesion and hence the results of molecular forces, without the need to resolve simulations at the molecular scale.

Similar to the modified collision operator for walls, we need to apply a modified force at solid nodes to account for adhesion at the fluid-solid interfaces. There exist many variations in the literature, e.g.\ Refs.~\cite{Huang2007,Li2019,Sedahmed2022,Pereira2022}. Most of them have in common that they use a gradient of some macroscopic field scaled by some interaction strength, i.e.\ 
\begin{equation}
    \vec{F}^\mathrm{solid}_\kappa(t,\x) = - \rho_\kappa(t,\x) 
        \sum_{\sigma\in\{\mathrm{solids}\}} G^\mathrm{solid}_{\kappa\sigma} \vec{\nabla} s_{\sigma}(t,\x).
\label{eq:force_solid}
\end{equation}
Here, $G^\mathrm{solid}_{\kappa\sigma}$ is the interaction strength of fluid component $\kappa$ with solid material $\sigma$. The field $s_\sigma$ is somewhat arbitrary, since many approaches or interpretations lead to the same results. Let us define it is a solid indicator field that is independent of time, such that
\begin{equation}
    s_\sigma(\x) = \begin{cases}
    1,\qquad\text{if solid material $\sigma$ is present at $\x$}\\
    0,\qquad\text{otherwise}.
    \end{cases}
\end{equation}
Note, that a common variant is to prescribe a fictitious density inside the solid nodes to compute a density gradient just like for the fluid-fluid interaction, cf.~\cite{Krueger_2017:LBM_Book}. We use the indicator field approach, because the fictitious density variant introduces an unnecessary additional parameter, which can be absorbed into the interaction strength. Depending on the choice of parameters, the two variants are equivalent.

Throughout this work, we apply Shan-Chen forcing \cite{ShanChen1993,ShanDoolen1995,MartysChen1996}, which modifies macroscopic velocities instead of including an explicit force term into \Cref{eq:LBM}. This is often done for simplicity or ease of implementation.

The scheme modifies the macroscopic velocity of component $\kappa$, such that
\begin{equation}
    \u_\kappa = \tilde{\u} + \frac{\tau_\kappa \vec{F}_\kappa}{\rho_\kappa},
\end{equation}
where $\tilde{\u}$ is referred to as common velocity, the relaxation time-weighted average velocity of all components, i.e.\ 
\begin{equation}
    \tilde{\u} = \frac{\sum_\kappa \rho_\kappa \u_\kappa / \tau_\kappa}{\sum_\kappa \rho_\kappa / \tau_\kappa}.
\end{equation}
All other variables are defined as above but refer to the $\kappa$-indexed fluid component. An additional external force field $\vec{F}_\kappa^\mathrm{ext}$, for example gravity, can be incorporated by replacing
\begin{equation}
    \vec{F}_\kappa \rightarrow \vec{F}_\kappa + \vec{F}_\kappa^\mathrm{ext}.
\end{equation}

In simulations, we will only have setups with two components, A and B, and a single solid material, which simplifies the discussion.

One should note that \Cref{eq:force_fluid} is in LBM literature usually not written using a gradient, cf.\ e.g.\ Ref.~\cite{Krueger_2017:LBM_Book}. Although the original sources still mention the relation \cite{ShanChen1993,ShanDoolen1995} a formulation using the lattice weights has become the dominant approximation in the LBM literature which we refer to as classical gradient operator in sections below. When not explicitly written as a gradient, it is not easy to be recognized as such, because the expression summing over lattice weights reminds more of the Gauss-Hermite quadrature to discretize integrals.\footnote{%
For the LBM discretization formalism compare Ref.~\cite{Kellers2026}.
} Some authors picked up the gradient formulation again, e.g.\ Ref.\ \cite{Peng2006}, and Shan even suggested a more accurate computation of the term similar to a finite difference expansion \cite{Shan2006,Shan2008}.

\subsection{Disadvantages of the Classical Gradient Operator}

In the following, we want to investigate properties of the model, especially concerning the force fields in \Cref{eq:force_fluid,eq:force_solid}. Most often in the context of LBM, those expressions are written in a more explicit lattice form. For example, \Cref{eq:force_fluid,eq:force_solid} can be found in the form
\begin{equation}
    \vec{F}_\kappa(t,\x) = - \rho_\kappa(t,\x) 
        \sum_{\kappa'} G_{\kappa\kappa'} \sum_i w_i \rho_{\kappa'}(t,\x+\c_i\Delta t)\c_i\Delta t,
\label{eq:force_fluid_PP}
\end{equation}
and
\begin{equation}
    \vec{F}^\mathrm{solid}_\kappa(t,\x) = - \rho_\kappa(t,\x) 
        \sum_{\sigma\in\{\mathrm{solids}\}} G^\mathrm{solid}_{\kappa\sigma} \sum_i w_i s(\x+\c_i\Delta t)\c_i\Delta t,
\label{eq:force_solid_PP}
\end{equation}
where $w_i$ is the weight for direction $i$. This form somewhat obfuscates the fact that we are computing discrete spatial gradients. Additionally, there is confusion about the weights, since it is often stated that they are the lattice weights, e.g.\ in Ref.\ \cite{Krueger_2017:LBM_Book}. While using the lattice weights is possible, it does not necessarily result in a valid gradient. We will prove this in \Cref{sec:methodology}. The confusion might stem from the original works \cite{ShanChen1993,ShanDoolen1995}, when LBM was still commonly used on hexagonal lattices where the weights are unity, such that the weights had to be included into the model only at a later point. In fact, Shan and Chen never claimed that lattice weights were required and Shan even clarified the meaning of the weights and the gradient formulation in later publications \cite{Shan2006,Shan2008}.

% In the following we want to highlight two related properties that work to our disadvantage and show the resulting errors in simulations. The first will be a lack of rotational symmetry of the resulting gradient fields. The second will be the related varying surface normal length.

%%%%%%%%%%%%%%%%%%%%%%%%%%%%%%%%%%%%%%%%%%%%%%%%%%%%%%%%%%%%%%%%%%%%%%%%%%%%%%%
\begin{figure}
    \centering
    \includegraphics[width=0.5\linewidth]{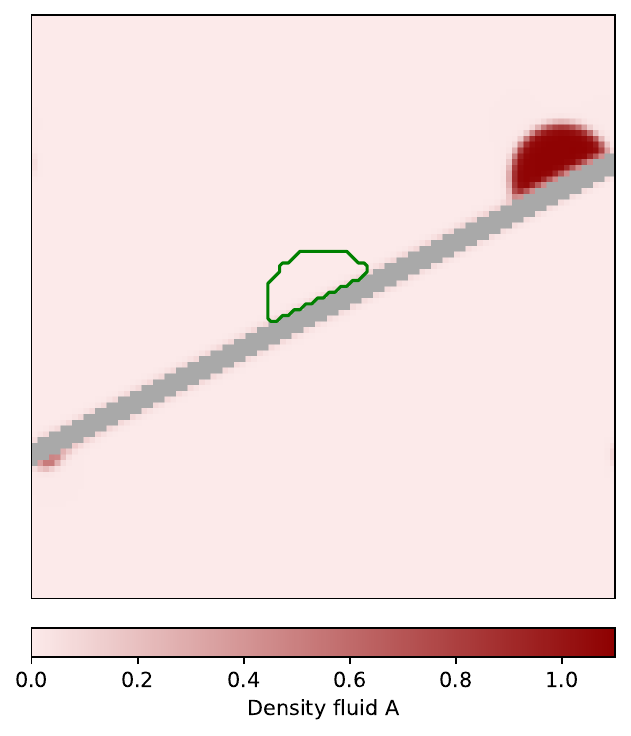}
    \caption{Droplet on a wall with a 50\% slope. Shown is the density field for fluid A at $t=5000\Delta t$. The green contour marks the initial position at $t=0$. Details in text.}
    \label{fig:moving_droplet}
\end{figure}
%%%%%%%%%%%%%%%%%%%%%%%%%%%%%%%%%%%%%%%%%%%%%%%%%%%%%%%%%%%%%%%%%%%%%%%%%%%%%%%

When using lattice weights as gradient weights, the gradient in \Cref{eq:force_fluid_PP} does come with a property that makes it undesirable for the computation of density gradients in the pseudopotential model: it is not rotationally symmetric. Therefore, the model works well in the bulk fluid and even with walls aligned with the lattice directions. However, simulations start to predict unphysical behavior once one introduces solid that breaks the symmetry of the lattice. As a result, walls that are not aligned with any of lattice directions lead to spurious currents in the fluid. Droplets on slanted walls, for example, will move because of these currents even without including gravitational forces in the simulations. This behavior is shown in \Cref{fig:moving_droplet}. Here, we simulated a $128\times128$ domain with two fluid components, A and B. A droplet is put on a wall (gray voxels), which is at a 50\% slope. This corresponds almost to a 22.5° angle, which would be maximally unaligned with the velocity directions given by $\c_i$. The density field of component A it shown for $t = 5000 \Delta t$. The green contour marks the initial position of the droplet. While this situation should physically be completely stable in absence of external forces, the droplet clearly moves on its own. The effect is small because we simulated a neutral wall with $\vec{F}^\mathrm{solid}_\mathrm{A,B} = \vec{0}$. Inclusion of adhesive forces would drastically increase the effect because they compound the symmetry breaking by increasing the strength of the forces towards walls.

The aforementioned adhesive forces highlight another problem with the model which is also related to rotational symmetry. A gradient operator of the form of \Cref{eq:force_solid_PP} applied to a solid indicator field results in gradient vector which is an approximation of a surface normal. Surface normals are usually normalized to unity. In voxelized structures the normalization needs to change depending on the relative position of the considered voxels because the distance to the wall changes. A rotationally symmetric operator includes this property, while the classical LBM gradient will produce differing contact angles on slanted walls.

In the next section, we will rewrite the model into a convolution-based gradient scheme, highlighting some of its properties. This will also allow for an easy introduction of the maximally rotationally symmetric variant of the discrete gradient operator. Gradients based in this operator will be properly normalized by design, so that we can avoid a manual directional correction of the surface normals.

% Throughout this work, $s(\x)$ is assumed to be constant in time. Therefore it is possible to precompute the surface normals or solid gradients. There exist many methods to do so. To avoid introduction of an additional degree of freedom in our comparisons by selecting a specific one, we showcase in the following only setups where the solid boundary can be described by an analytical expression.

% In the literature, there exist many variations of \Cref{eq:force_solid_PP}, for example replacing the binary solid indicator field with a fictitious density inside walls.\todo{\cite{}} This does not change the fact that in the end we are always just estimating surface normals of the solid. 

\section{Methodology}
\label{sec:methodology}

\subsection{Discrete Gradient Operators as Convolutions}

The task at hand given by \Cref{eq:force_fluid,eq:force_solid} is to compute discrete gradients of some discrete fields, $\rho(\x)$ and $s(\x)$. This is a known problem in many different fields, and hence many approaches to solve this problems are formulated in different disciplines. A particularly elegant and efficient approached was developed in the field of image analysis. The approach is based on the idea of applying a gradient kernel $g_\ia$ for spatial direction $\ia$ to $\rho_\kappa$ via a convolution. With that, we can write \Cref{eq:force_fluid} as
\begin{equation}
    \vec{F}_\kappa(t,\x) 
    % = - \rho_\kappa(t,\x) \sum_{\kappa'} G_{\kappa\kappa'} \nabla \rho_\kappa'(\x)
    = - \rho_\kappa(t,\x) \sum_{\kappa'} G_{\kappa\kappa'} \sum_{\ia=1}^d 
        \left[g_\ia \ast \rho_\kappa'(t,
        \x)\right] \hat{e}_\ia,
\label{eq:force_fluid_convolution}
\end{equation}
and \Cref{eq:force_solid} analogously as
\begin{equation}
    \vec{F}^\mathrm{solid}_\kappa(t,\x) 
    % = - \rho_\kappa(t,\x) 
    %     \sum_{\sigma\in\{\mathrm{solids}\}} G^\mathrm{solid}_{\kappa\sigma} \vec{\nabla} s_{\sigma}(t,\x)
    = - \rho_\kappa(t,\x) \sum_{\sigma\in\{\mathrm{solids}\}} G^\mathrm{solid}_{\kappa\sigma} \sum_{\ia=1}^d 
        \left[
            g_\ia \ast s_{\sigma}(t,\x) 
        \right] \hat{e}_\ia.
\label{eq:force_solid_convolution}
\end{equation}
We introduced $\hat{e}_\ia$, the unit vector of spatial direction $\ia$ in $d$ dimensions and the binary operator $\ast$ for the discrete convolution. It is defined in \Cref{app:A}.

The pseudopotential approaches using the classical gradient operator can actually be transferred to this convolution notation. In this form the notation hints at some of its relevant properties. For example, assuming a D2Q9 LBM lattice with standard weights, \Cref{eq:force_fluid_convolution} with kernels
\begin{equation}
    g_x = \frac{1}{36}
    \begin{bmatrix}
    -1 & 0 & 1\\
    -4 & 0 & 4\\
    -1 & 0 & 1
    \end{bmatrix},\qquad
    g_y = \frac{1}{36}
    \begin{bmatrix}
     1 & 4 & 1\\
     0 & 0 & 0\\
    -1 &-4 &-1
    \end{bmatrix},
\label{eq:filters}
\end{equation}
is equivalent to \Cref{eq:force_fluid_PP}, if $w_i$ are indeed the weights of a D2Q9 lattice. A proof is given in \Cref{app:A}.

One property of higher dimensional convolution kernels is that they can be separated into convolutions of lower dimensional ones. Taking again the 2D kernels $g_{x,y}$ as an example, we can write them as convolutions of two 1D kernels, such that
\begin{equation}
    g_x = 
    \frac{1}{36}
    \begin{bmatrix}
    -1 & 0 & 1\\
    -4 & 0 & 4\\
    -1 & 0 & 1
    \end{bmatrix}
    =
    \frac{1}{36}
    \begin{bmatrix}
    -1 & 0 & 1
    \end{bmatrix}
    \ast 
    \begin{bmatrix}
    1 \\ 4 \\ 1
    \end{bmatrix}
    ,
\end{equation}
\begin{equation}
    g_y = 
    \frac{1}{36}
    \begin{bmatrix}
     1 &  4 &  1\\
     0 &  0 &  0\\
    -1 & -4 & -1
    \end{bmatrix}
    =
    \frac{1}{36}
    \begin{bmatrix}
    4 & 1 & 4
    \end{bmatrix}
    \ast 
    \begin{bmatrix}
    1 \\ 0 \\ -1
    \end{bmatrix}
    .
\end{equation}
From this we recognize that gradient kernels are always a combination of a central difference kernel and a smoothing kernel with a suitable prefactor for normalization. The prefactor is arbitrary in the pseudopotential force, since any change of normalization can be absorbed into the interaction parameters $G$. Generally, we find that gradient kernels of the form
\begin{equation}
    k_x = 
    \begin{bmatrix}
    -a & 0 & a\\
    -b & 0 & b\\
    -a & 0 & a
    \end{bmatrix},
    \qquad
    k_y = 
    \begin{bmatrix}
     a &  b &  a\\
     0 &  0 &  0\\
    -a & -b & -a
    \end{bmatrix}
\end{equation}
can always be factored as
\begin{equation}
    k_x = 
    \begin{bmatrix}
    -a & 0 & a\\
    -b & 0 & b\\
    -a & 0 & a
    \end{bmatrix}
    =
    \begin{bmatrix}
    -1 & 0 & 1
    \end{bmatrix}
    \ast 
    \begin{bmatrix}
    a \\ b \\ a
    \end{bmatrix},
    \qquad
    k_y = 
    \begin{bmatrix}
     a &  b &  a\\
     0 &  0 &  0\\
    -a & -b & -a
    \end{bmatrix}
    =
    \begin{bmatrix}
    a & b & a
    \end{bmatrix}
    \ast 
    \begin{bmatrix}
    1 \\ 0 \\ -1
    \end{bmatrix}.
\end{equation}
which can be interpreted as the weights $w_i$ only affecting the smoothing part of the gradient. For reasonable numerical values of $w_i$ the resulting gradient varies only slightly, but it might have different properties. Those properties directly influence the quality of LBM multicomponent simulations.

The factorization extends to 3D by using a 1D finite difference kernel $\mathcal{C}$ in the gradient direction, and two 1D smoothing kernels $\mathcal{D}$ corresponding for the other two directions. Using tensors and index notation, we can write this as
\begin{equation}
    \begin{split}
        \mathcal{K}_{x,\ia\ib\ic} 
        = \mathcal{C}_\ia \mathcal{D}_\ib \mathcal{D}_\ic,
        \\
        \mathcal{K}_{y,\ia\ib\ic} 
        = \mathcal{D}_\ia \mathcal{C}_\ib \mathcal{D}_\ic,
        \\
        \mathcal{K}_{z,\ia\ib\ic} 
        = \mathcal{D}_\ia \mathcal{D}_\ib \mathcal{C}_\ic,
    \end{split}
\end{equation}
resulting in convolution kernels $\mathcal{K}_i$ for each direction, which are rank 3 tensors. The direction of the gradient $i$ is determined by the position of the central difference kernel in the tensor product. 

The properties of gradient convolution kernels are very general and hold for 3D, as well as extended lattices where we would encounter kernels larger than $3\times 3$ or $3\times 3\times 3$.

\subsection{Maximally Rotational Symmetric Convolution Kernel: Scharr Operator}

In image analysis, gradient operators are often used as edge detectors. For most weights, the gradient kernels come with the same property as the interaction force model above: the resulting gradients are not rotationally symmetric. 
It is, however, possible to construct a type of maximally rotational symmetric gradient operator.\footnote{It is important to note that there is no single one maximally rotationally symmetric operator. The reason for this is that one has freedom of choice in selecting the error metric when comparing against the ideal analytical case. Additionally, this error will never be exactly zero for all rotations}
Scharr \cite{Scharr2000} derived it by minimizing the error of the 22.5° case, which is maximally unaligned with the lattice velocities in Fourier space. This is done by comparison with the Fourier transform of the continuous gradient operator. In its standard normalization, the resulting $3\times 3$ operators in 2D read
\begin{equation}
    \mathcal{S}_x = 
    \begin{bmatrix}
    -46.84 & 0 & 46.84 \\
    -162.32 & 0 & 162.32 \\
    -46.84 & 0 & 46.84
    \end{bmatrix},
    \qquad
    \mathcal{S}_y = 
    \begin{bmatrix}
    46.84 & 162.32 & 46.84 \\
    0 & 0 & 0 \\
    -46.84 & -162.32 & -46.84
    \end{bmatrix}.
\label{eq:Scharr_operator}
\end{equation}
We will refer to the kernels as Scharr operators in the following.

\subsection{Simulation Setups}
\label{subsec:setups}

In the following, we will always simulate two fluid components, $\kappa \in \{\mathrm{A},\mathrm{B}\}$, interacting with a single solid, such that we can drop the material index $\sigma$.

The gradient kernels are normalized. The finite difference part with $1/2$, and the smoothing part such that the total operator conserves the field quantity acted upon, i.e.\ 
\begin{equation}
    k_x =
    \frac{1}{2}
    \begin{bmatrix}
    -1 & 0 & 1
    \end{bmatrix}
    \ast 
    \frac{1}{2a+b}
    \begin{bmatrix}
    a \\ b \\ a
    \end{bmatrix}
    =
    \frac{1}{4a+2b}
    \begin{bmatrix}
    -a & 0 & a\\
    -b & 0 & b\\
    -a & 0 & a
    \end{bmatrix},
\end{equation}
\begin{equation}
    k_y =
    \frac{1}{2a+b}
    \begin{bmatrix}
    a & b & a
    \end{bmatrix}
    \ast 
    \frac{1}{2}
    \begin{bmatrix}
    1 \\ 0 \\ -1
    \end{bmatrix}
    =
    \frac{1}{4a+2b}
    \begin{bmatrix}
     a &  b &  a\\
     0 &  0 &  0\\
    -a & -b & -a
    \end{bmatrix}.
\end{equation}
This allows for stable simulations using fluid-fluid interaction of unity, 
\begin{equation}
    G_{\mathrm{AB}} = 1.
\end{equation}
We already mentioned above that a multicomponent model does not require self interaction of the fluids, such that
\begin{equation}
    G_{\mathrm{AA}} = G_{\mathrm{BB}} = 0.
\end{equation}

We apply the implementation variant that separates fluid and solid interactions. Using $G_{\mathrm{AB}} = 1$, in 2D it corresponds to
\begin{equation}
    \vec{F}_\mathrm{A}(t,\x) 
    = - \rho_\mathrm{A}(t,\x) \sum_{\ia\in\{x,y\}} 
        \Bigl[
            k_\ia \ast 
                (1-s(\x)) \rho_\mathrm{B}(t,\x) 
        \Bigr]
        \hat{e}_\ia,
\end{equation}
and
\begin{equation}
    \vec{F}^\mathrm{solid}_\mathrm{A}(t,\x) 
    = - \rho_\mathrm{A}(t,\x) 
        G_\mathrm{A}^\mathrm{solid}
        \sum_{\ia\in\{x,y\}} 
        \Bigl[
            k_\ia \ast s(\x) 
         \Bigr]
        \hat{e}_\ia,
\end{equation}
and analogously for fluid B with the labels A and B swapped. The kernels $k_\ia$ are either the classical ones, $g_\ia$, or the Scharr operators, $\mathcal{S}_\ia$.

We will also use two variations of the solid interaction: "symmetric" and "B only". In the symmetric case we use
\begin{equation}
    G_\mathrm{B}^\mathrm{solid} = -G_\mathrm{A}^\mathrm{solid} = \hat{G},
\end{equation}
where $\hat{G}$ is a constant that has to be calibrated to result in the desired contact angle.

For the case that applies solid interaction to fluid B only, we set
\begin{equation}
    G_\mathrm{A}^\mathrm{solid} = 0, \qquad G_\mathrm{B }^\mathrm{solid} = \tilde{G}.
\end{equation}
This approach creates no gradients for fluid A on fluid-solid interfaces. The constant $\tilde{G}$ will be chosen such that we get a comparable contact angle as for the symmetric case.

\section{Results}
\label{sec:results}

\subsection{Contact Angles from Droplets on Walls}
\label{subsec:contact_angles}

%%%%%%%%%%%%%%%%%%%%%%%%%%%%%%%%%%%%%%%%%%%%%%%%%%%%%%%%%%%%%%%%%%%%%%%%%%%%%%%
\begin{figure}
    \centering
    \includegraphics[width=0.7\linewidth]{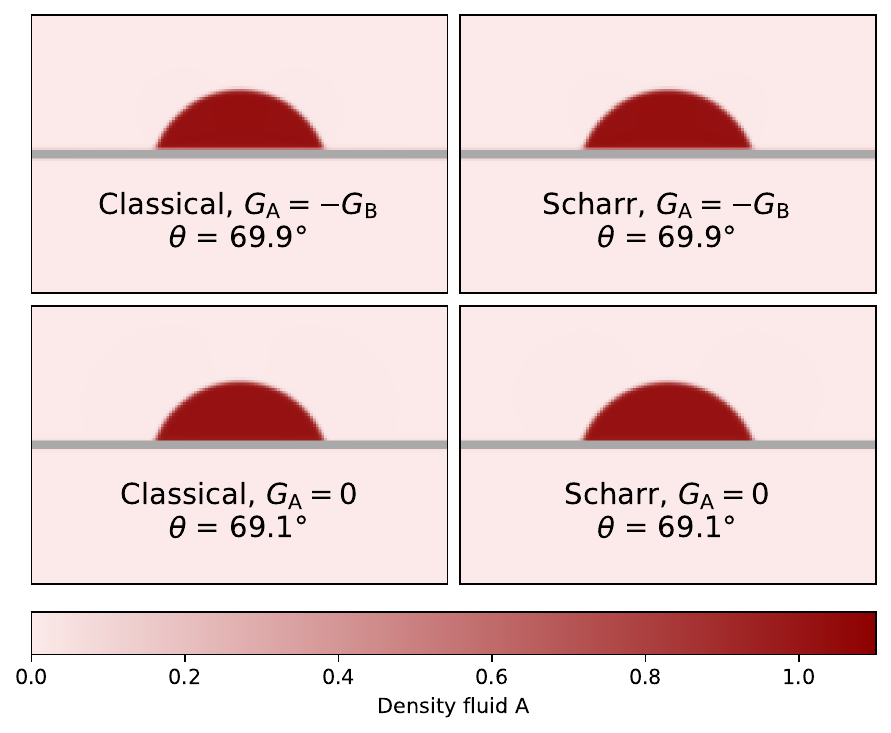}
    \caption{Droplet tests of the different approaches to measure contact angles. The angles were measured at $t=15'000\Delta t$. Left column: the classical LBM gradient operator using lattice weights. Right column: Gradient computed with Scharr operators. Top row: symmetric solid interaction. Bottom row: Solid interaction applied to fluid B only.}
\label{fig:calibration}
\end{figure}
%%%%%%%%%%%%%%%%%%%%%%%%%%%%%%%%%%%%%%%%%%%%%%%%%%%%%%%%%%%%%%%%%%%%%%%%%%%%%%%

First, we calibrate the setups by finding suitable values for $\hat{G}$ and $\tilde{G}$. For later setups we prefer a wetting contact angle $\theta < 90\deg$. The contact angle is measured by putting droplets on flat solid surfaces. Without gravity, the droplets will form perfectly circular disk segments. Hence, we can compute the contact angle $\theta$ via
\begin{equation}
    \tan \theta = \frac{C}{2(R-S)},
\end{equation}
where $S$ is the height of the droplet (sagitta) and $C$ the length of the covered wall section (chord length). The radius of the corresponding disk $R$ can be computed via
\begin{equation}
    \frac{4S^2 + C^2}{8S}.
\end{equation}
To measure $S$, we count all voxels where $\rho_\mathrm{A} > 0.5$ as part of the droplet. For $C$ we do it similarly, but consider that the symmetric solid interaction creates a density gradient at the wall, such that we have to set a lower threshold of $\rho_\mathrm{A} > 0.2$.

For the symmetric case, we simply set $\hat{G} = 0.20$ and measure the resulting angle. The value for $\tilde{G}$ that sets an equivalent contact angle, which was found to be $\tilde{G} = 0.18$. The large deviation from the Young equation (\Cref{eq:Young}) for the latter case is a known phenomenon and has been investigated by Huang et al.\ in Ref.~\cite{Huang2007}.

The resulting contact angle measurements for the four combinations of gradient kernels and wall interaction variants (cf.\ \Cref{subsec:setups}) are shown in \Cref{fig:calibration}. The systems have been equilibrated for $15'000$ time steps before measuring the angles. The initial density in bulk regions is set to be $1.0$, while the dissolved density in the surrounding regions is $0.001$. The droplet was initialized as a half disk with $\theta=90\deg$, which has its center point horizontally centered at the wall. All boundaries are periodic.

We do not interpolate for the computation of $S$ and $C$, but only consider full voxels. Therefore, the simulations will not be able to produce arbitrary contact angles, and the four angles in \Cref{fig:calibration} are equivalent within the range of measurement uncertainty.

\subsection{Surface Tension Measurements}
\label{subsec:surface_tension}

%%%%%%%%%%%%%%%%%%%%%%%%%%%%%%%%%%%%%%%%%%%%%%%%%%%%%%%%%%%%%%%%%%%%%%%%%%%%%%%
\begin{figure}
    \centering
    \includegraphics[width=0.6\linewidth]{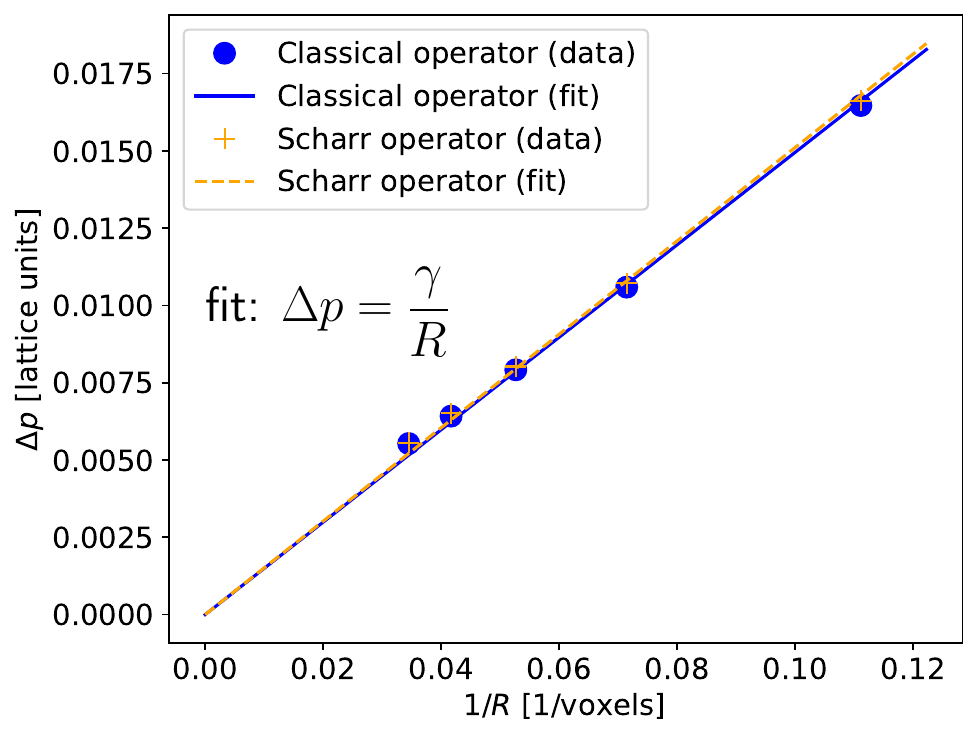}
    \caption{Surface tension measurement for the classical and Scharr operator sets using five different radii in a bubble test.}
\label{fig:surface_tension}
\end{figure}
%%%%%%%%%%%%%%%%%%%%%%%%%%%%%%%%%%%%%%%%%%%%%%%%%%%%%%%%%%%%%%%%%%%%%%%%%%%%%%%

For the analysis of capillary rise, we require the surface tension for the chosen values of $\tilde{G}$ and $\hat{G}$. It can be measured from a bubble test through an application of the Young-Laplace equation. In 2D, it reads
\begin{equation}
    \Delta p = \frac{\gamma}{R},
\label{eq:YoungLaplace}
\end{equation}
where $\Delta p$ is the pressure difference between the inside and the outside of a bubble, $\gamma$ is the surface tension, and $R$ is the bubble radius.

The radius is again measured by considering voxels with $\rho_\mathrm{A} > 0.5$ as belonging to the bubble because the initial bulk density was again 1.0. We initialized five different values for $R$, and measure the resulting $\Delta p$ after 5000 time steps. In a two-component system consisting of fluid A and B, pressure is computed via
\begin{equation}
    p(\x) = \cs^2 \bigl[ \rho_\mathrm{A}(\x) + \rho_\mathrm{B}(\x) \bigr]
        + \cs^2 G_\mathrm{AB} \rho_\mathrm{A}(\x) \rho_\mathrm{B}(\x),
\end{equation}
with $\cs=1/3$ being the lattice speed of sound. Because of the normalization of the gradient kernels, the fluid-fluid interaction strength is unity, i.e.\ $G_\mathrm{AB} = 1$.

\Cref{fig:surface_tension} shows the fit for the surface tension measuring the $\Delta p$ for 5 different radii. The results in lattice units for both gradient operator sets are
\begin{align*}
    &\text{Classical}: &&\gamma = 0.149 \pm 0.002 
    \qquad\qquad \qquad\qquad \qquad\qquad \\
    &\text{Scharr}:    &&\gamma = 0.151 \pm 0.002
    \qquad\qquad \qquad\qquad \qquad\qquad 
\end{align*}
which are not significantly different. The given uncertainty is the square root of the variance of the estimated parameter, i.e.\ the standard deviation, which can be computed from the covariance matrix of the fit routine. In our case it is \texttt{optimize.curve\_fit} of the Python library SciPy \cite{SciPy}.

\subsection{Symmetry Breaking Wall Orientation}
\label{subsec:wall}

%%%%%%%%%%%%%%%%%%%%%%%%%%%%%%%%%%%%%%%%%%%%%%%%%%%%%%%%%%%%%%%%%%%%%%%%%%%%%%%
\begin{figure}
    \centering
    \includegraphics[width=0.99\linewidth]{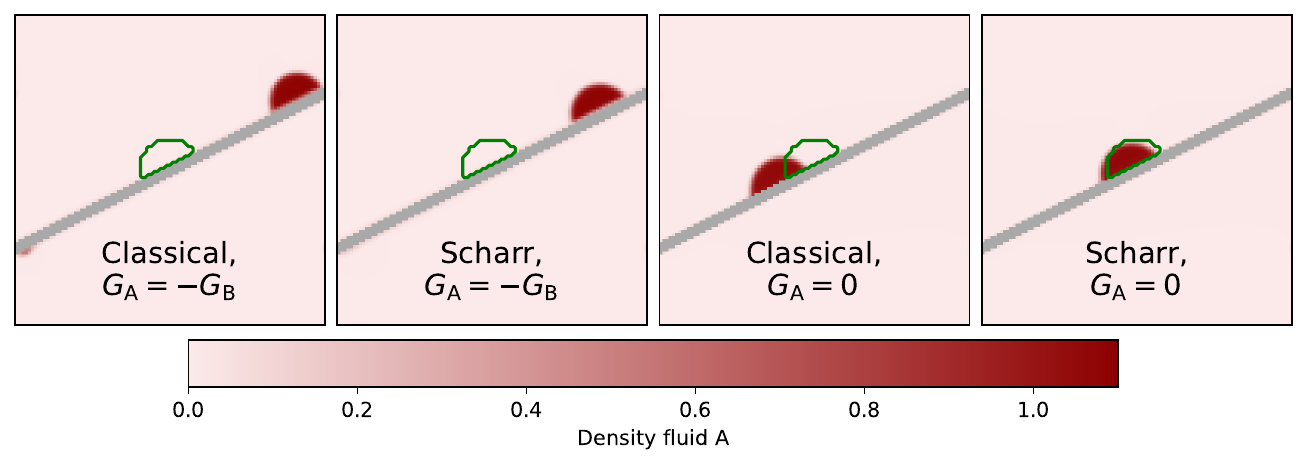}
    \caption{Droplets on slanted walls at $t=1500\Delta t$. The green outlines show the initial droplet positions. Left: Classical operator set with symmetric solid interaction. Center left: Scharr operators with symmetric solid interaction. Center Right: Classical operators with solid interaction through fluid B only. Right: Scharr operators with solid interaction through fluid B only. Details in text.}
\label{fig:wall}
\end{figure}
%%%%%%%%%%%%%%%%%%%%%%%%%%%%%%%%%%%%%%%%%%%%%%%%%%%%%%%%%%%%%%%%%%%%%%%%%%%%%%%

Next, we demonstrate how the application of the Scharr operator set and a non-symmetric wall interaction changes the simulation predictions for droplets on slanted walls. The setup is again a droplet initially centered on a slanted wall with a $50\%$ slope, which is closest to the maximally unaligned case of $22.5\deg$ without introduction of rounding error in the voxelized solid. The other simulation parameters are as above.

The resulting droplet positions after 1500 time steps are shown in \Cref{fig:wall}. The image shows the situation for the classical operators and for the Scharr operators, each using both variants of solid interaction.

It is somewhat expected that the Scharr operator performs better than the classical approach using lattice weights. The symmetry breaking of the unaligned solid causes the gradient computation to introduce spurious currents. But the results also show that the Scharr operator alone cannot fully negate the introduction of fictitious velocities. However, combined with the approach were all solid interaction comes through fluid B, the droplet position is stable within the simulation time. The small variation from the initial contour stems from the fact that the initial fluid distribution with a 90\textdegree{} contact angle is not a stable configuration. Equilibration in initial time steps slightly adjusts its fluid distribution.

\subsection{Jurin's Law}
\label{subsec:jurin}

%%%%%%%%%%%%%%%%%%%%%%%%%%%%%%%%%%%%%%%%%%%%%%%%%%%%%%%%%%%%%%%%%%%%%%%%%%%%%%%
\begin{figure}
    \centering
    \includegraphics[width=0.8\linewidth]{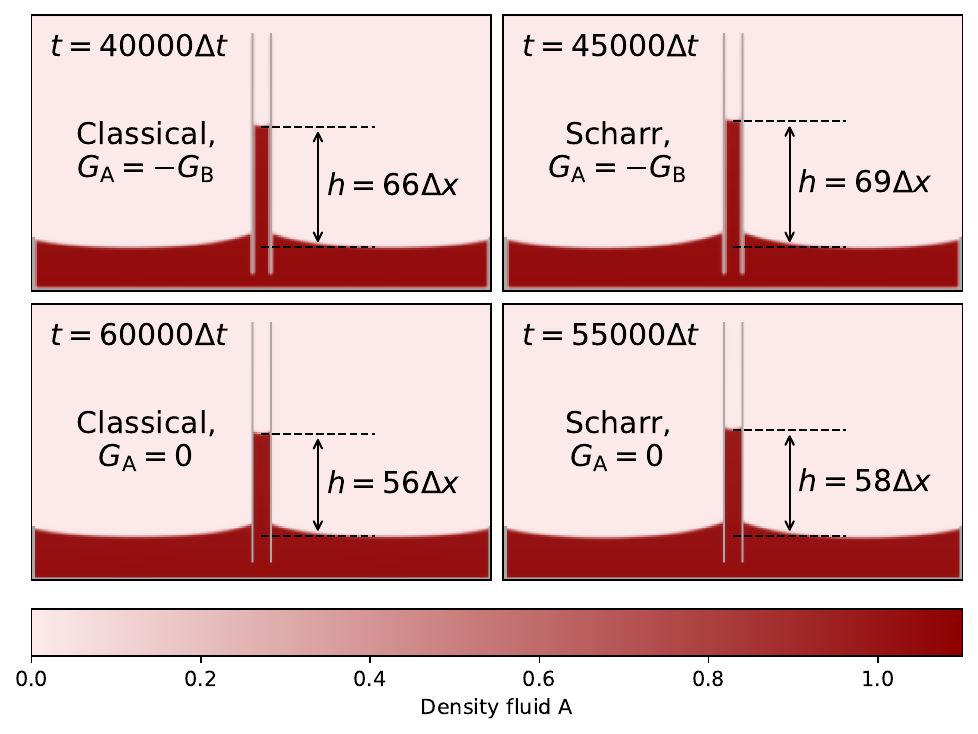}
    \caption{Final capillary heights in a Jurin's Law test comparing the classical operator set (left) and the Scharr operator set (right). Both were combined with the symmetric solid interaction (top) and with interaction introduced through fluid B only (bottom). Solid walls are shown in gray. Details in text.}
\label{fig:jurin}
\end{figure}
%%%%%%%%%%%%%%%%%%%%%%%%%%%%%%%%%%%%%%%%%%%%%%%%%%%%%%%%%%%%%%%%%%%%%%%%%%%%%%%

Since a slanted flat wall is a somewhat artificial setup. We are more interested in quantifying the differences that would be present in more realistic simulations. To have access to a ground truth, we want to compare the model to a configuration that has an analytical solution. A good candidate is Jurin's law. It states that a wetting fluid rising in a capillary under the influence of gravity approaches a final height $h$ as $t \to \infty$. For fluid A rising in a 2D capillary that is aligned with the gravitational field, the height is given by
\begin{equation}
    h = \frac{\gamma \cos\theta}{g \rho_\mathrm{A}^\mathrm{bulk} r_\mathrm{c}},
\end{equation}
where $\gamma$ is the surface tension, $\theta$ the contact angle, $g$ the acceleration due to gravity, $\rho_\mathrm{A}^\mathrm{bulk}$ the bulk density of fluid A, and $r_\mathrm{c}$ the radius of the capillary. 

The simulations are done with $g=0.0002$ voxels/$\Delta t^2$ in the negative $y$-direction, and a capillary diameter $2r_\mathrm{c} = 9$ voxels. From \Cref{subsec:surface_tension} we assume the surface tension to be $\gamma = 0.15$ in lattice units. Considering small deviations in $\rho_\mathrm{A}^\mathrm{bulk}$ and the different contact angles we measured in \Cref{subsec:contact_angles}, the average value for the expected height comes out to $\bar{h} \approx 58 \text{ voxels}$.

In \Cref{fig:jurin} we compare both operator sets and each combined with the two solid interaction variants. Gravity acts on fluid A only to simulate a system similar to water/air, were the gravitational acceleration can be neglected for the much lighter fluid B. The fluid height in the capillary and the basin were initially identical. Simulations are stopped when the system reaches a stationary state identified by no change in height for 5000 time steps. Therefore, the overall simulation time varies in the different cases. The rest of the parameters were chosen as above.

Our simulations indicate that the Jurin's law setup does not benefit from a rationally symmetric gradient operator, because the solid is ideally aligned with lattice directions. But it greatly benefits from the modification in the fluid interaction. If all solid interaction is introduced through fluid B, fluid A has no gradient at fluid-solid interfaces. A gradient will effectively narrow the capillary, especially when it is very thin. This leads to an overestimation of the capillary rise as the above results clearly show. This is why our results for fluid interaction through fluid B only are overall better while the symmetric case overestimates the height. In fact, a model with Scharr operators and gradient-free walls leads to perfect agreement with the expected height.

\subsection{Rotational Symmetry for Porous Media Intrusion}
\label{subsec:porous}

%%%%%%%%%%%%%%%%%%%%%%%%%%%%%%%%%%%%%%%%%%%%%%%%%%%%%%%%%%%%%%%%%%%%%%%%%%%%%%%
\begin{figure}
    \centering
    \includegraphics[width=0.9\linewidth]{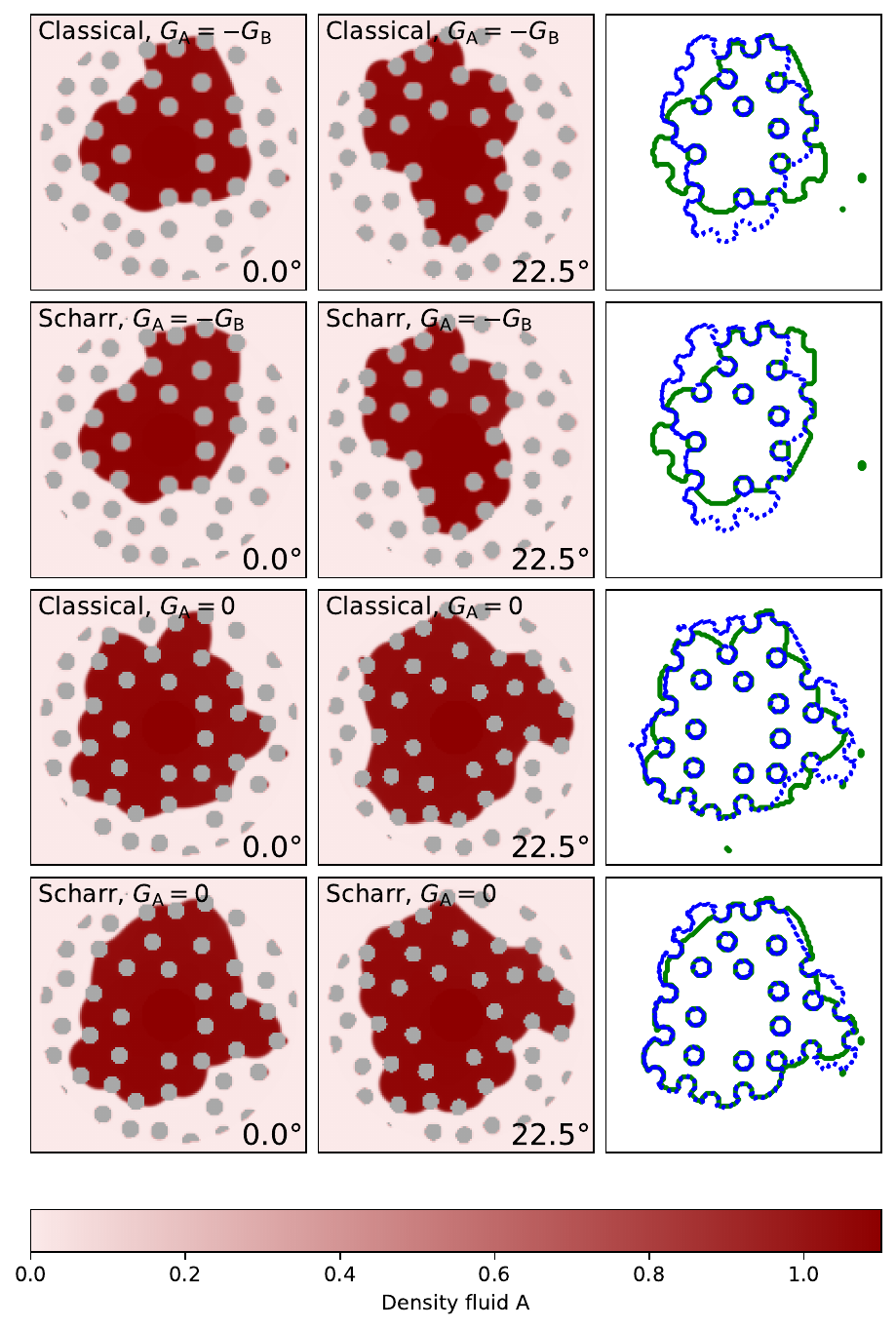}
    \caption{Rotational symmetry test in a porous media application. The left column used the unchanged medium, while in the central column it was rotated counter-clockwise by $22.5\deg$ to have the rotation maximally unaligned to D2Q9 lattice directions. From top to bottom: Classical operators with symmetric interaction, Scharr operators with symmetric interaction, classical operators with interaction through fluid B, and Scharr operators with interaction through fluid B. The right columns show the contours of the invaded regions, where the blue dotted line is rotated back clockwise to compare the overlap with the unrotated case shown in solid green. All simulations ran for 7500 time steps.}
\label{fig:porous}
\end{figure}
%%%%%%%%%%%%%%%%%%%%%%%%%%%%%%%%%%%%%%%%%%%%%%%%%%%%%%%%%%%%%%%%%%%%%%%%%%%%%%%

To test the model in a more realistic setting, we compare how the two interaction variants respectively combined with the two gradient operator sets perform when simulating multicomponent fluid flow in a porous medium. The simulation domain is filled with non-overlapping solid disks. Their center points are generated randomly and used to create two digital versions representing the same medium. For the first one, we use the unchanged center coordinates. In the second one, the center coordinates are rotated by $22.5\deg$ counter-clockwise before generating the image. The domain measures $151\times151$ voxels to have one voxel exactly at the center, which improves the symmetry when working with radial boundary conditions. The random disks all have a radius of 4.5 voxels---all voxels with a smaller Euclidean distance from the corresponding center point are turned solid. The relatively small size was chosen on purpose to avoid smooth solid walls because we intended to mimic low resolution scans of a real porous medium.

To check for rotational symmetry, the boundary conditions also need to be as radially symmetric as possible, otherwise the two different rotations would not correspond to the same physical system. Inlet and outlet both have fixed density and zero velocity boundary conditions for the present phase, and we arbitrarily chose $\rho_\mathrm{A}^\mathrm{inlet} = 1.1$ and $\rho_\mathrm{B}^\mathrm{outlet} = 1.0$ to create a positive pressure difference. To not introduce mass exchange through the dissolved densities, fluid A and B experience bounce-back at the outlet and inlet, respectively. The inlet is a concentric disk with radius 15, while the outlet is every voxel that falls outside a concentric disk of radius 70. This introduces a radial pressure gradient that drives the invasion of fluid A into the material. To make the outlet more visually apparent and to avoid errors in the outlet boundary condition, we removed the solid in this region. All other simulation parameters are the same as above.

The results are shown in \Cref{fig:porous}. We compare the classical and the Scharr operator sets, each applied with either symmetric wall interaction or solid interaction through fluid B only. For visual comparison of the invaded regions, the third column shows the contours for $\rho_\mathrm{A}=0.5$, where the one for the rotated image has been rotated back to create overlap.

Our results show partly expected and partly unexpected behavior. It is clear that a gradient operator that is optimized for rotational symmetry should perform better in this example. It is also somewhat expected that in narrow channels we get similar behavior as in Jurin's law, meaning the capillary pressure of a solid throat might be overestimated leading to differing invasion patterns. What was not expected is the fact that the symmetric solid interaction does in fact perform better compared to the case where only fluid B interacts with the wall and $G_\mathrm{A}=0$. Considering that the latter performed much better in the slanted wall setup (\Cref{fig:moving_droplet}), this is somewhat unexpected. However, it is also the case that a symmetric interaction force combination has more potential to even out numerical errors, which becomes more relevant in real porous media applications compared to the somewhat artificial test setups shown above.

\section{Discussion and Outlook}
\label{sec:discussion}

In our paper we present improvements to the well established and broadly applied pseudopotential methods for multiphase flow. The pseudopotential model is known for causing spurious currents, but the effect becomes increasingly large with symmetry breaking solids, which is unavoidable in porous media applications. In our paper we introduce two strategies to mitigate typical issues in multiphase flow simulations including interactions with solid walls.

We demonstrate that the wetting behavior in simulations can be modeled more accurately by introducing a rotationally symmetric gradient operator to compute the interaction forces. By adopting techniques developed in the image processing community this comes with zero additional computational costs compared to the classical method. 

In two-component systems the pseudopotential method can benefit from introducing solid interaction only through one of the two fluids. However, this strongly depends on the simulation setup. Like in the shown example for Jurin's law, we are often interested in the behavior of one fluid, and the second fluid can be considered an interaction field to model the first one accurately. In those cases, we can delegate the solid interaction fully to the second fluid, and avoid numerical artifacts like density gradients at walls for the main fluid. Otherwise those gradients would slightly modify contact angles or result in effectively smaller throat diameters when invading porous media or capillaries.

By introducing radially symmetric pressure boundary conditions, we were able to benchmark the presented model variants against each other, where the Scharr operator always performed better than the classical gradient operator based on LBM lattice weights. The choice of the solid interaction model depends on the simulation setup and simulated processes.

Future work will include the application of the 3D implementation of this model to more realistic inhomogeneous and anisotropic porous media. Especially the radially symmetric pressure boundary condition presented here might give new insights when extracting properties like pressure-saturation relationships for those materials. For demonstration purposes, we showed here only relatively short simulations that ran for a few thousand time steps. It will be interesting to investigate, if the deviations between the two operator sets remain roughly constant over time or if they increase more and more in cases where the fluid has enough time to spread out differently across the domain.

\paragraph*{Acknowledgements:}
The authors wish to extend their sincere thanks to Julius Weinmiller for useful discussion.
The presented research contributes to the Center for Electrochemical Energy Storage Ulm-Karlsruhe (CELEST).

\paragraph*{Author Contributions:}
Conceptualization, methodology, software, validation, formal analysis, investigation, resources, data curation, visualization, and writing (original draft): BK.
Writing (review\,\&\,editing): BK, AL, TD.
Supervision, project administration, and funding acquisition: AL, TD.
All authors confirm that they have read the final manuscript and agree to its publication.

\paragraph*{Conflicts of Interest:}
The authors declare no conflicts of interest.

\section*{Abbreviations}
We used the following abbreviations in the text:

\begin{tabular}{ll}
    LBM &   Lattice-Boltzmann method \\
    BGK &   Bhatnagar–Gross–Krook 
\end{tabular}

\printbibliography

\appendix
% \appendix\setcounter{section}{1}
\crefalias{subsection}{appendix}
\crefalias{section}{appendix}

\section{Appendix}

\subsection{Proof of Classical Method and Convolution Approach Equivalence}
\label{app:A}

In units where $\Delta t = 1$, the D2Q9 LBM lattice is characterized by the velocity vector set
\begin{equation}
    \{\c_0,...,\c_8 \} = 
    \left\lbrace 
        \begin{pmatrix} 0 \\ 0 \end{pmatrix},
        \begin{pmatrix} 1 \\ 0 \end{pmatrix},
        \begin{pmatrix} 0 \\ 1 \end{pmatrix},
        \begin{pmatrix} -1 \\ 0 \end{pmatrix},
        \begin{pmatrix} 0 \\ -1 \end{pmatrix},
        \begin{pmatrix} 1 \\ 1 \end{pmatrix},
        \begin{pmatrix} -1 \\ -1 \end{pmatrix},
        \begin{pmatrix} 1 \\ -1 \end{pmatrix},
        \begin{pmatrix} -1 \\ 1 \end{pmatrix}
    \right\rbrace,
\end{equation}
and the corresponding set of weights
\begin{equation}
    \{w_0,...,w_8\} = 
    \left\lbrace 
        \frac{16}{36},
        \frac{4}{36},
        \frac{4}{36},
        \frac{4}{36},
        \frac{4}{36},
        \frac{1}{36},
        \frac{1}{36},
        \frac{1}{36},
        \frac{1}{36}
    \right\rbrace.
\end{equation}
For a compact notation, we define
\begin{equation}
    \x_i = \x + \c_i
\end{equation}

We compute the relevant sum of \Cref{eq:force_fluid_PP} for an arbitrary discrete field $\phi(\x)$, i.e.\ 

\begin{equation}    
\begin{split}
    \sum_{i=1}^8 w_i \phi(\x+\c_i)\c_i
    = \phantom{+} &\frac{4}{36} 
    \left[ 
        \phi(\x_1)
            \begin{pmatrix} 1 \\ 0 \end{pmatrix}
        + \phi(\x_2)
            \begin{pmatrix} 0 \\ 1 \end{pmatrix}
        + \phi(\x_3)
            \begin{pmatrix} -1 \\ 0 \end{pmatrix}
        + \phi(\x_4)
            \begin{pmatrix} 0 \\ -1 \end{pmatrix}
    \right]
    \\ 
    + &\frac{1}{36} 
    \left[ 
        + \phi(\x_5)
            \begin{pmatrix} 1 \\ 1 \end{pmatrix}
        + \phi(\x_6)
            \begin{pmatrix} -1 \\ -1 \end{pmatrix}
        + \phi(\x_7)
            \begin{pmatrix} 1 \\ -1 \end{pmatrix}
        + \phi(\x_8)
            \begin{pmatrix} -1 \\ 1 \end{pmatrix}
    \right].
\end{split}
\end{equation}
After factoring this into basis vector components, we arrive at
\begin{equation}
\begin{split}
    \sum_{i=1}^8 w_i &\phi(\x+\c_i)\c_i \\
    = \phantom{+}
    &\frac{1}{36}
    \Bigl[
          4\phi(\x_1)
        - 4\phi(\x_3)
        + \phi(\x_5)
        - \phi(\x_6)
        + \phi(\x_7)
        - \phi(\x_8)
    \Bigr]
    \begin{pmatrix} 1 \\ 0 \end{pmatrix}
    \\
    + &\frac{1}{36}
    \Bigl[
          4 \phi(\x_2)
        - 4 \phi(\x_4)
        + \phi(\x_5)
        - \phi(\x_6)
        - \phi(\x_7)
        + \phi(\x_8)
    \Bigr]
    \begin{pmatrix} 0 \\ 1 \end{pmatrix}
\end{split}
\end{equation}

The 2D convolution kernels correspoding to the classic pseudopotential method using the lattice weights of D2Q9 are
\begin{equation}
    g_x = 
    \frac{1}{36}
    \begin{bmatrix}
    -1 & 0 & 1\\
    -4 & 0 & 4\\
    -1 & 0 & 1
    \end{bmatrix}
    ,\qquad
    g_y = 
    \frac{1}{36}
    \begin{bmatrix}
     1 &  4 &  1\\
     0 &  0 &  0\\
    -1 & -4 & -1
    \end{bmatrix}
    .
\end{equation}
It is helpful to define a displacement vector matrix
\begin{equation}
    \mathcal{D} =  
    \begin{bmatrix}
     \c_8  &  \c_2 &  \c_5\\
     \c_3  &  \c_0 &  \c_1\\
     \c_6  &  \c_4 &  \c_7
    \end{bmatrix}
\end{equation}
making it easier to do a convolution by hand. For an arbitrary kernel $K$, the discrete convolution with an arbitrary field $\phi(\x)$ is
\begin{equation}
    K \ast \phi(\x) = 
        \sum_{\ia,\ib} K_{\ia\ib} \cdot \phi(\x + \mathcal{D}_{\ia\ib}).
\end{equation}

The discrete gradient in \Cref{eq:force_fluid_convolution} for $\phi(\x)$ would be
\begin{equation}
\begin{split}
    \nabla \phi(\x) &= 
        \Bigl[g_x \ast \phi(\x)\Bigr] \begin{pmatrix} 1 \\ 0 \end{pmatrix}
        + \Bigl[g_y \ast \phi(\x)\Bigr] \begin{pmatrix} 0 \\ 1 \end{pmatrix}
        \\
        &=\phantom{+}
        \left\lbrace
        \frac{1}{36}
        \begin{bmatrix}
        -1 & 0 & 1\\
        -4 & 0 & 4\\
        -1 & 0 & 1
        \end{bmatrix}
        \ast
        \phi(\x)
        \right\rbrace
        \begin{pmatrix} 1 \\ 0 \end{pmatrix}
        \phantom{=}+ 
        \left\lbrace
        \frac{1}{36}
        \begin{bmatrix}
         1 &  4 &  1\\
         0 &  0 &  0\\
        -1 & -4 & -1
        \end{bmatrix}
        \ast
        \phi(\x)
        \right\rbrace
        \begin{pmatrix} 0 \\ 1 \end{pmatrix}
        \\
        &=\phantom{+}
        \frac{1}{36}
        \Bigl[
              4\phi(\x+\c_1)
            - 4\phi(\x+\c_3)
            + \phi(\x+\c_5)
            - \phi(\x+\c_6)
            + \phi(\x+\c_7)
            - \phi(\x+\c_8)
        \Bigr]
        \begin{pmatrix} 1 \\ 0 \end{pmatrix}
        \\
        &\phantom{=}+
        \frac{1}{36}
    \Bigl[
          4 \phi(\x+\c_2)
        - 4 \phi(\x+\c_4)
        + \phi(\x+\c_5)
        - \phi(\x+\c_6)
        - \phi(\x+\c_7)
        + \phi(\x+\c_8)
    \Bigr]
    \begin{pmatrix} 0 \\ 1 \end{pmatrix}
    \\
        &=\phantom{+}
        \frac{1}{36}
        \Bigl[
              4\phi(\x_1)
            - 4\phi(\x_3)
            + \phi(\x_5)
            - \phi(\x_6)
            + \phi(\x_7)
            - \phi(\x_8)
        \Bigr]
        \begin{pmatrix} 1 \\ 0 \end{pmatrix}
        \\
        &\phantom{=}+
        \frac{1}{36}
    \Bigl[
          4 \phi(\x_2)
        - 4 \phi(\x_4)
        + \phi(\x_5)
        - \phi(\x_6)
        - \phi(\x_7)
        + \phi(\x_8)
    \Bigr]
    \begin{pmatrix} 0 \\ 1 \end{pmatrix}.
\end{split}
\end{equation}
We therefore conclude
\begin{equation}
    \sum_{i=1}^8 w_i \phi(\x+\c_i)\c_i
    =
    \Bigl[g_x \ast \phi(\x)\Bigr] \begin{pmatrix} 1 \\ 0 \end{pmatrix}
    + \Bigl[g_y \ast \phi(\x)\Bigr] \begin{pmatrix} 0 \\ 1 \end{pmatrix}
    =
    \nabla \phi(\x).
\end{equation}

\subsection{Implementation and Parametrization Details of Pseudopotential Methods}
\label{app:B}

\subsubsection*
{Force Components}

In specific implementation, one is faced with treating some freedom of choice given by \Cref{eq:force_fluid_PP,eq:force_solid_PP}. In the literature, there exist multiple variations on how to treat solid material in neighboring lattice sites, i.e.\ at positions $\x+\c_i\Delta t$. One approach is to let the sum in \Cref{eq:force_fluid_PP} run over all neighboring nodes. If the node is solid, the expression is computed with a fictitious value for the density inside the solid. This value is often chosen to be constant, but there are also approaches that try to tweak the behavior of the model with variable ``solid densities''. As a result, any gradient that would be yielded by \Cref{eq:force_solid_PP} can be set with a specific value for this solid density and be directly included into \Cref{eq:force_fluid_PP}. This leads to a huge number of approaches in the literature, which are often equivalent or subsets of a more general theory, compare Refs.\ \cite{Huang2007,Li2019,Sedahmed2022,Pereira2022}.

In another variant, the solid neighboring nodes are excluded from \Cref{eq:force_fluid_PP}. This is in good approximation equivalent to a solid density of zero, neglecting dissolved density effects. The solid-fluid interfaces are then fully treated by \Cref{eq:force_solid_PP}. Fluid-fluid and fluid-solid interfaces are that way somewhat decoupled and can be treated using differing approaches.

\subsubsection*{Interaction Strengths}

The choice of interaction strengths also introduces freedom of choice. Usually, $G_{\kappa\kappa'}$ is somewhat fixed. Its value is physically related to the surface tension, while it numerically determines the width of the diffuse interfaces. If it is too small, and the interfaces will diffuse out and no stable fluid bubbles will form. Set too large, the simulations will become unstable, since diffuse interface models with too narrow interfaces lead to arbitrarily high gradients. It can be avoided by proper normalization of the gradient operators resulting in a model in which $G_{\kappa\kappa'}$ = 1 yields stable interfaces.

Contact angles are a result of the different surface tensions of different interfaces. For a contact angle $\theta_\sigma$ on solid $\sigma$ they are related through the Young equation
\begin{equation}
    \cos \theta_\sigma = \frac{
        \chi_{\kappa'\sigma}
        - \chi_{\kappa\sigma}
    }{\gamma_{\kappa\kappa'}},
\label{eq:Young}
\end{equation}
where $\gamma_{\kappa\kappa'}$ is the surface tension of components $\kappa$ and $\kappa'$, and $\chi_{\kappa\sigma}$ is the surface tension between fluid component $\kappa$ and solid material $\sigma$. Similar to experiments, where values for $\chi_{\kappa\sigma}$ and $\chi_{\kappa'\sigma}$ can not be measured independently, the contact angle in simulations is determined only by the difference $G_{\kappa'\sigma}^\mathrm{solid}-G_{\kappa\sigma}^\mathrm{solid}$ rather than the specific values of $G_{\kappa\sigma}^\mathrm{solid}$ and $G_{\kappa'\sigma}^\mathrm{solid}$.

\end{document}